\documentclass[prb,reprint]{revtex4-1}
\usepackage{blindtext}
\usepackage{siunitx} % Provides the \SI{}{} and \si{} command for typesetting SI units
\usepackage{graphicx} % Required for the inclusion of images
\usepackage{natbib} % Required to change bibliography style to APA
\usepackage[utf8]{inputenc}
\usepackage{braket}
\usepackage[table]{xcolor}
\xdefinecolor{RWTHBlue}{RGB}{0,84,159}
\xdefinecolor{myRWTHBlue}{RGB}{64,127,183}
\usepackage[
	colorlinks=true,
  	urlcolor=RWTHBlue,	
  	linkcolor=RWTHBlue,
  	citecolor=RWTHBlue
  	]{hyperref}
\usepackage{units}
\usepackage{amssymb}
\usepackage[toc,page]{appendix}
\usepackage[english]{babel}
\usepackage{amsmath} % Required for some math elements
\usepackage{gensymb} % Required for degree symbel

  \DeclareGraphicsExtensions{.pdf}
\usepackage{import}

\newcommand{\rfig}[1]{Fig.\,\ref{#1}}

\newcommand{\req}[1]{Eq.\,(\ref{#1})}
\newcommand{\rsec}[1]{Sec.\,\ref{#1}}

\usepackage{xspace}

\newcommand{\AlO}{Al$_{2}$O$_{3}$\xspace}

\newcommand{\sysS}{$S$\xspace}
\newcommand{\sysM}{$M$\xspace}
\newcommand{\sysL}{$L$\xspace}
\newcommand{\toymodel}{toy-model\xspace}

\begin{document}
\title{Calculation of tunnel-couplings  in open gate-defined disordered quantum dot systems}
\author{Jan Klos}
\affiliation{JARA-FIT Institute Quantum Information, Forschungszentrum	J\"ulich GmbH and RWTH Aachen University, D 52074 Aachen, Germany}
\author{Fabian Hassler}
\affiliation{JARA-FIT Institute Quantum Information, Forschungszentrum	J\"ulich GmbH and RWTH Aachen University, D 52074 Aachen, Germany}
\author{Pascal Cerfontaine}
\affiliation{JARA-FIT Institute Quantum Information, Forschungszentrum	J\"ulich GmbH and RWTH Aachen University, D 52074 Aachen, Germany}
\author{Hendrik Bluhm}
\affiliation{JARA-FIT Institute	Quantum Information, Forschungszentrum	J\"ulich GmbH and RWTH Aachen University, D 52074 Aachen, Germany}
\author{Lars R. Schreiber}
\email{schreiber@physik.rwth-aachen.de}
\affiliation{JARA-FIT Institute Quantum Information, Forschungszentrum	J\"ulich GmbH and RWTH Aachen University, D 52074 Aachen, Germany}

\begin{abstract}
Quantum computation based on semiconductor electron-spin qubits requires high control of tunnel-couplings, both across quantum dots and between the quantum dot and the reservoir. The tunnel-coupling to the reservoir sets the qubit detection and initialization bandwidth for energy-resolved spin-to-charge conversion and is essential to tune single-electron transistors commonly used as charge detectors. Potential disorder and the increasing complexity of the two-dimensional gate-defined quantum computing devices sets high demands on the gate design and the voltage tuning of the tunnel barriers. We present a Green's formalism approach for the calculation of tunnel-couplings between a quantum dot and a reservoir. Our method takes into account in full detail the two-dimensional electrostatic potential of the quantum dot, the tunnel barrier and reservoir. A Markov approximation is only employed far away from the tunnel barrier region where the density of states is sufficiently large. We calculate the tunnel-coupling including potential disorder effects, which become increasingly important for large-scale silicon-based spin-qubit devices. Studying the tunnel-couplings of a single-electron transistor in Si/SiGe as a showcase, we find that charged defects are the dominant source of disorder leading to variations in the tunnel-coupling of four orders of magnitude.
\end{abstract}

\maketitle
\section*{Introduction}
Gate-defined quantum dots (QDs) have proved to be a versatile platform for confining charge, electron-spin and hole-spin quantum bits (qubits) in various material systems. Tremendous progress has been achieved in planar AlGaAs \cite{Hansona, Nowack2011a, Shulman2012, Medford2013, Yoneda2014, Botzem2015} and Si-based systems \cite{Zwanenburg2013} such as CMOS structures \cite{Veldhorst2014b, Veldhorst2015b}, SiGe \cite{Borselli2014, Zajac2016, Watson2017, Yoneda2017} and Si nanowires \cite{Voisin2014, Betz2016, Brauns2016, Maurand2016}. Focusing on scalability towards large-scale quantum systems \cite{Hill2015, Vandersypen2016, Veldhorst2016}, the complexity of the gate design increases, trending to denser gate configurations of QDs \cite{Delbecq2014,Baart2016,Borselli2014,Zajac2015,Zajac2016,Flentje2017}.
For scaling towards large numbers of qubits, it is essential to design the electrostatic gate patterns such that key parameters are nearly equal for each qubit, despite the typical electrostatic disorder present, due to imperfections of the host crystal lattice. Examples of such parameters are the inter-QD tunnel-coupling and QD-to-reservoir tunnel-coupling. Specifically, the tunnel coupling from QD to electron reservoir has to be well controlled for spin-to-charge conversion schemes involving spin-state dependent tunneling\cite{Elzermann2004,Hansona}.
Charge read-out of multiple QDs in close proximity has been demonstrated using single electron transistors (SET), for which tunnel barriers to both source and drain reservoirs have to be properly set\cite{Barthel2009}.
Tunnel-couplings can be tuned by gate-voltages over a wide range \cite{Borselli2011,Rochette2017}. Automatic tuning of a large number of quantum dots\cite{Kalantre2017} would require however that the tunnel couplings can be calculated for disorder potentials. Optimizing the gate design in this respect requires taking the details of the potential in the vicinity of the tunnel barriers into account.
The increasing complexity of large-scale devices makes gate design development based on iterative fabricational and experimental studies alone very inefficient. Specific properties such as electrostatic disorder can be simulated prior to sample fabrication \cite{Frees2016}.\newline
The tunnel-coupling between two QDs (closed system) can be numerically calculated by solving the Schrödinger equation. Calculating the tunnel-coupling between a QD and a reservoir (here defined as open system) solving the full system is challenging. Several different approaches to take the tunnel coupling between a QD and a reservoir into account have been used e.g. master equation based \cite{Ziegler2000} or a transfer Hamiltonian \cite{Bardeen1961, Noguera1989}. Prominent is the Wentzel-Kramers-Brillouin (WKB) approximation, which is based on a semi-classical, one dimensional trajectory of an electron \cite{Garg2000,Jelic2012,Das2014}.\newline
In this work, we present an approach for calculating the tunnel coupling in an open two-dimensional system based on Green's formalism with the Markov approximation. 
% Inserted by Frederica
%
Applying the Markov approximation only far away from the tunnel barrier, this approach allows to capture potential details of the reservoir region in close proximity of the QD. The calculation of the tunnel coupling is exact in principle and can be adapted to available computational resources by setting a boundary within the two-dimensional reservoir. The boundary divides the potential region which is fully quantum mechanically captured from the shapeless Markov-approximated region. 
We validated our method on a two-dimensional model system with $N$ sites and find the analytically calculated value for the tunnel coupling within a $6 \%$ error. 
The remaining small discrepancy is a result of our used tight-binding model. The resulting error in the tunnel coupling could be easily compensated by tuning gate voltages during an experiment.
We apply our method of calculating the tunnel couplings on an SET in a Si/SiGe heterostructure as a showcase. Since our method captures full details of the electrostatic potential, we are able to study the effect of three different types of electrostatic disorder sources considered to be present in Si/SiGe heterostructures.
For our SET gate design, we find that charged defects at the heterostructure surface are dominant and can lead to variation in the tunnel coupling of four orders of magnitude. \newline
This paper is structured as follows:
In \rsec{Theory}, we present the method for the calculation of tunnel couplings in open quantum systems based on Green's formalism.
In \rsec{Application}, we use the presented method on the electrostatic potential landscapes of our SET gate design including three different disorder effects present in a Si/SiGe heterostructure as a showcase. 
In \rsec{Verification}, our numerical method is applied on an analytic 2D toy-model system as a benchmark test.
\begin{figure}[htb]
\begin{center}

  		\includegraphics[width=0.48\textwidth]{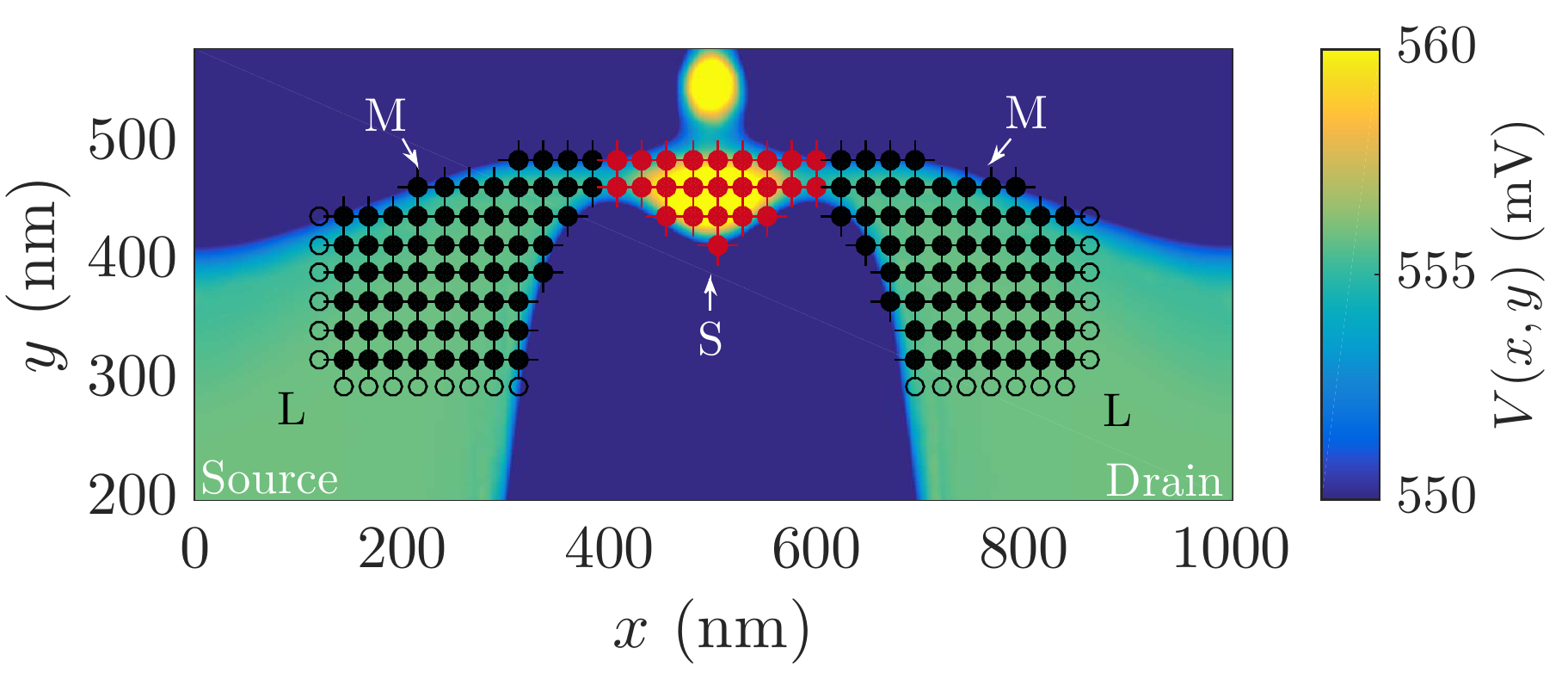}
\end{center}
\caption{Electrostatic potential of the SET overlayed by 2D tight-binding model using nearest-neighbor coupling coupled to a quantum dot in a Si/SiGe heterostructure. The whole system (read-out QD and reservoir) is divided into three subsystems containing the read-out QD (system \sysS depicted in red), the electronic reservoir far away from the read-out QD (system \sysL depicted by black circles) and an intermediate system (system \sysM depicted by black dots) connecting \sysS and \sysL. Using the Markov approximation, only the first sites of system \sysL have to be used. For the calculations, we consider a much higher density of sites (dots) than plotted here.\newline}
\label{fig:2DShowcase}
\end{figure}
\section{Theory}\label{Theory}
%This section focuses on the general analytic derivation of the approach. Without further restrictions, the derivation is based on a total system containing three adjacent, non-overlapping subsystems labeled \sysS, \sysM and \sysL. To give a meaning to these subsystems, \sysS corresponds to a closed system such as a QD, \sysM corresponds to an intermediate system such as the tunnel barrier and the adjacent part of a reservoir and \sysL corresponds to the rest of the reservoir also called lead system. System \sysS and \sysM are defined as closed systems whereas \sysL leads to decay. All systems are coupled to their adjacent systems, specifically \sysS to \sysM and \sysM to \sysL. The tunnel coupling between a QD and reservoir, system \sysS and $M+L$ respectively, corresponds to the decay of the states of the electron within the QD based on the coupled lead system.\newline
For calculating the tunnel coupling $t_\mathrm{C}$ between a QD and its reservoirs, we take the following approach. We divide the whole system (dot and reservoir), into three adjacent, non-overlapping subsystems: system \sysS, which represents the QD; system \sysL, which represents the electronic reservoir far away from the QD; and system \sysM which is an intermediate region connecting \sysS and \sysL (see \rfig{fig:2DShowcase}).  Each sub-system is tunnel coupled to the neighboring one. We are interested in the level broadening  of the eigenstates of \sysS due to the coupling to \sysM+\sysL. In a tunnel-Hamiltonian description in which system \sysS is directly coupled to the reservoir, this level broadening is directly related to the tunnel coupling matrix element $t_\mathrm{C}$ between QD and reservoir.\newline 
We will treat system \sysL in the wide-band Markov approximation, meaning that we assume an energy independent constant density of states $\rho_L$. Physically, this corresponds to assuming that system \sysL is not affected by the system \sysS + \sysM, and that all electrons injected into \sysL cannot return to the system.\newline
For the calculation of $t_\mathrm{C}$, we follow a Greens formalism approach analog to Ref. \cite{Datta1995}. For the lead system with Hamiltonian $H_\mathrm{L}$, the Greens function operator is defined by \begin{equation}\label{eq:Grho}
\hat{G}_\mathrm{L}(\hbar\omega) = \dfrac{1}{\hbar\omega - H_\mathrm{L}} \ \mathrm{.}
\end{equation}
where $\hbar \omega$ is the energy parameter and $\hbar$ is the reduced Planck constant. Rewriting the operator $\hat{G}_\mathrm{L}(\hbar\omega)$ in its eigenbasis, we get the scalar Greens function $G_\mathrm{L}(\hbar\omega)$. Using the Kramers-Kronig relation, the scalar Greens function is derived using the corresponding density of states $\rho_\mathrm{L}$ of the leads with
\begin{equation} \label{eq:Grscalar}
G_\mathrm{L}(\hbar\omega)  = \int \dfrac{\mathrm{d \omega'}}{2\pi\hbar} \dfrac{\rho_\mathrm{L}(\omega')}{\omega - \omega' + i \eta^{+}} \ \mathrm{,}
\end{equation}
where $\eta^{+}$ is a positive regularization factor.
Since the actual density of states of the reservoir is unknown, we assume a wide-band Markov approximation with constant $\rho_\mathrm{L}(\hbar\omega)$ \cite{CohenTamoudji1989}. Hence, \req{eq:Grho} simplifies to
\begin{equation}\label{eq:Markov}
G_\mathrm{L}(\hbar\omega) = -i\pi \rho_\mathrm{L}
\end{equation}
Alternatively, Green's formalism is capable to describe the reservoir system analytically by infinite 2D plane waves. This leads to additional challenges e.g. choosing a suitable 2D representations of plane waves, which are out of scope of this work. 
Focusing on subsystem \sysM coupled to the lead system and integrating out the lead, the effective Hamiltonian of the reservoir is 
\begin{equation}\label{eq:Hmeff}
H_\mathrm{M,eff} = H_\mathrm{M} + w^{\dagger}_\mathrm{ML} \hat{G}_\mathrm{L} w_\mathrm{ML}
\end{equation}
with the Hamiltonian $H_\mathrm{M}$ of the isolated intermediate system and  $w_\mathrm{ML}$ the coupling matrix between \sysM and \sysL. $H_\mathrm{M,eff}$ is diagonalized with the eigenvalues $\epsilon_\mathrm{m}-i\gamma_\mathrm{m}$ with $\gamma_\mathrm{m}>0$ and left eigenvectors $\langle \Psi_\mathrm{m'} |$ and right eigenvectors $| \Psi_\mathrm{m} \rangle$. Note that $\langle \Psi_\mathrm{m'} | \neq | \Psi_\mathrm{m} \rangle^\dagger$ since $H_\mathrm{M,eff}$ is non-hermitian but both eigenvectors fulfill the bi-orthogonality relation
\begin{equation}
\langle \Psi_\mathrm{m'} | \Psi_\mathrm{m} \rangle = \delta_\mathrm{m'm} \ .
\end{equation} 
With this procedure, we find the Green's function operator of subsystem \sysM to be
\begin{equation}\label{eq:Gmeff}
\hat{G}_\mathrm{M}(\hbar\omega) 
= \dfrac{1}{\hbar\omega - H_\mathrm{M,eff}} 
= \sum_\mathrm{m} 
\dfrac{ | \Psi_\mathrm{m} \rangle \langle \Psi_\mathrm{m} | }{\hbar\omega -\epsilon_\mathrm{m}+i\gamma_\mathrm{m}+i\gamma_\mathrm{ext}} \ \mathrm{,}
\end{equation}
where we introduce $\gamma_\mathrm{ext}$ as an additional external regularization parameter which compensated for the finite number of sites numerically taken into account. In section \rsec{Verification}, we discuss the optimization of $\gamma_\mathrm{ext}$ in details. Focusing on subsystem \sysS, the Hamiltonian $H_\mathrm{S}$  is solved by $H_\mathrm{S} |s\rangle = \epsilon_\mathrm{s}|s\rangle $ with the eigenvector $|s\rangle$ and its corresponding eigenvalue $\epsilon_\mathrm{s}$.
The time-evolution of a state $|\mathrm{s}\rangle $ is described by its retarded Greens function
\begin{equation}\label{eq:Gst}
G_\mathrm{S}(t) = -i \Theta(t) \langle \mathrm{s} | e^{ - i H_\mathrm{tot} t} | \mathrm{s} \rangle 
\end{equation}
with $H_\mathrm{tot}$ is the Hamiltonian of the total system in \sysS, where subsystem \sysS and \sysM are coupled by the matrix $w_\mathrm{SM}$ analog to \req{eq:Hmeff}.
The Fourier transform of \req{eq:Gst} is
\begin{equation}\label{eq:Gshw}
G_\mathrm{S}(\hbar \omega) = \dfrac{1}{\hbar \omega - \epsilon_\mathrm{s}-\Sigma_\mathrm{S}(\hbar \omega)} \ ,
\end{equation}
where
\begin{equation}\label{eq:Sigma}
\Sigma_\mathrm{S}(\hbar \omega)= \langle s | w^{\dagger}_\mathrm{SM} \hat{G}_\mathrm{M}(\hbar\omega)  w_\mathrm{SM}  | s \rangle
\end{equation}
is the self-energy. The real-part of $\Sigma_\mathrm{S}$ corresponds to an energetic shift within system \sysS induced by the coupled system \sysM also called Lamb-shift\cite{Lamb1940}. This Lamb-shift depends on all states within system \sysM. 
In the following, we assume weak-coupling between subsystem \sysS and $M+L$. This corresponds to the physical situation where $| \mathrm{s} \rangle$ is a well-defined state within \sysS.
The imaginary part of the self-energy $\Sigma_\mathrm{S}$ leads to an energy-level broadening in system \sysS, resembling a decay of the wavefunction $| s \rangle$. This decay corresponds to an electron within the QD, which tunnels via the intermediate system \sysM into the lead system. In this model, the energy-level broadening in \sysS corresponds to the tunnel coupling of the state $| \mathrm{s} \rangle$ given by 
\begin{equation}\label{eq:tC}
t_C = 2 \ \mathrm{Im}(\Sigma_\mathrm{S}(\epsilon_\mathrm{s})) \ \mathrm{,}
\end{equation}
where the factor 2 accounts for the decay of the probability instead of the probability amplitude as $|\Psi|^2 \propto \mathrm{exp}(-2 \mathrm{Im}(\Sigma_\mathrm{S})t/\hbar)= \mathrm{exp}(-t_\mathrm{C} t/\hbar)$.\newline
\subsubsection*{Implementation recipe}
For reference, we want to highlight all necessary steps to use the presented method for the calculation of tunnel-couplings.\newline
We start with a computed electrostatic potential containing QDs and electron reservoirs. The Thomas-Fermi approximation is used to describe electron reservoirs, which imply significant screening effects. Regions containing a QD are calculated using superposition of the induced electrostatic potential of the modeled gate design.
From this given electrostatic potential, the tunnel coupling is calculated by following three-step protocol:\newline
(I) We define the presented subsystems \sysS, \sysM. The truncation between \sysS and \sysM is defined perpendicular to the tunneling direction along the potential maximum of the tunnel barrier. At the maximum of the tunnel barrier the influence of the used boundary conditions is minimal for both subsystems. Subsystem \sysL is defined as the remaining part of the reservoir, which is not covered by \sysM and can be chosen by balancing out the importance of details of the reservoir potential versus computations power.\newline
(II) We define the corresponding Hamiltonians $H_\mathrm{S}$ and $H_\mathrm{M}$ and coupling matrices $w_\mathrm{SM}$ and $w_\mathrm{ML}$. 
Using \req{eq:Markov} with a constant 2D density of states and $w_\mathrm{ML}$ in \req{eq:Hmeff}, $H_\mathrm{M,eff}$ is defined.\newline
(III) By solving the eigenvalue problem of $H_\mathrm{S}$ and $H_\mathrm{M,eff}$, the self-energy $\Sigma_\mathrm{S}$ can be calculated using \req{eq:Gmeff} and \req{eq:Sigma}.
By solving the eigenvalue problem of $H_\mathrm{S}$ and inverting $G_\mathrm{M}(\hbar\omega)$, $\Sigma_\mathrm{S}$ can be calculated directly. Finally, the tunnel coupling $t_\mathrm{C}$ is calculated using \req{eq:tC}. Alternatively, $t_\mathrm{C}$ can also be calculated by using computational cheaper matrix inversion.%
\section{Tunnel-coupling in realistic systems}\label{Application}
\begin{figure*}[htb]
\begin{center}
	  \includegraphics[width=0.99\textwidth]{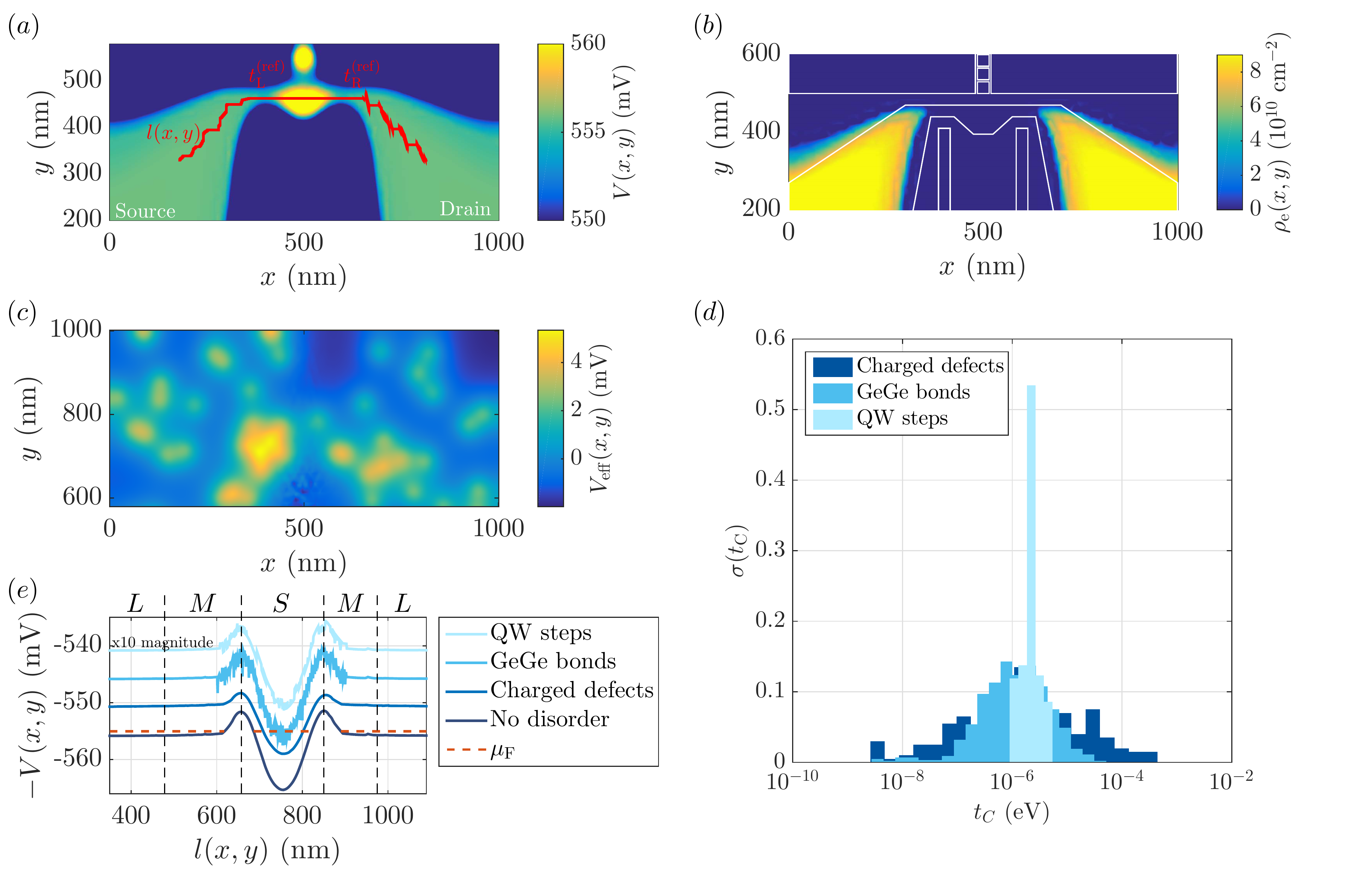}
\end{center}
\caption{
(a) Tuned electrostatic potential $V(x, y)$ forming two reservoirs (source, drain), one read-out QD and an adjacent QD within the 2DEG layer in a Si/SiGe heterostructure. A qualitative semi-classical tunnel path of an electron from source to drain is shown by $l(x,y)$.
(b) Corresponding electron density to (a) overlayed with the used gate structure (outlined by white lines). Thomas-Fermi approximation is used in regions of high electron density to include screening effects leading to a flat potential.
(c) Exemplary effective electrostatic potential induced by remote impurities located at the interface between the heterostructure and an oxide layer with a distance of $34 \ \mathrm{nm}$ to the QW and a positive charge $q_e=e$ with $e$ the electron charge.
(d) Normalized distribution of the simulated tunnel-couplings $\sigma(t_\mathrm{C})$ for different types of disorder.
QW steps as a possible source of disorder exhibit variations within one order of magnitude in $t_\mathrm{C}$ with $N_\mathrm{dis}=10^4$ random disorder configurations.
GeGe bonds as a possible source of disorder exhibit variations in two orders of magnitude around the reference value. 
Charged defects lead to variations in $t_\mathrm{C}$ of more than four orders of magnitude, $N_\mathrm{dis}=200$. The tunnel-couplings without any disorder are $t_\mathrm{L}^\mathrm{(ref)}=1.3 \ \mathrm{\mu eV}$ and $t_\mathrm{R}^\mathrm{(ref)}=2.0 \ \mathrm{\mu eV}$.
(e) The electrostatic potential evaluated along the semi-classical tunnel path $l(x,y)$ for three different types of disorder in comparison to the case of no disorder. Within the reservoir where the Thomas-Fermi approximation is used, the effects of the disorder are screened by electrons. Potentials are offset by $5 \ \mathrm{mV}$ for clarity. The potential fluctuations dues to QW steps has to be enlarged by a factor 10 prior to adding them to the gate-induced potential, because otherwise they are not visible in the plot. 
}

\label{fig:combinedApp}
\end{figure*}

In this section, we use the presented algorithm to calculate tunnel-couplings of an open system including potential disorder with three different length scales $\lambda$ in undoped Si/SiGe quantum wells. \newline
As a showcase, the electrostatic potential $V(x,y)$ of a QD capacitively coupled to a read-out QD of an SET is used and shown in \rfig{fig:combinedApp} (a) computed solving the 3D Poisson equation using COMSOL Multiphysics Software package\cite{comsol}. In regions of high electron concentrations e.g. reservoirs screening effects lead to flat electrostatic potentials. Here, the Thomas-Fermi approximation is used. The shape of these reservoirs is defined by potential barriers exceeding the Fermi energy $\mu_\mathrm{F}$. The resulting computed electron density is shown in \rfig{fig:combinedApp} (b). In regions of expected low electron concentrations e.g. QDs and tunnel barriers the electrostatic potential is calculated using a linear superposition of the electrostatic potential of every gate independently. Within this section the Fermi energy is defined by $\mu_\mathrm{F}=E_\mathrm{G}/2=555 \ \mathrm{meV}$ with the energy bandgap of silicon $E_\mathrm{G}=1.11 \ \mathrm{eV}$.\newline
We define our used tight-binding system using nearest-neighbor coupling with a spatial resolution $a=1 \ \mathrm{nm}$. The on-site potential $V_\mathrm{ij}$ is given by the previously computed electrostatic potential $V(x_\mathrm{i},y_\mathrm{j})$ at position $x_\mathrm{i} = x/a$ and $y_\mathrm{i}$ respectively. The nearest-neighbor coupling element is defined by $t_\mathrm{ij}=\hbar^2 \Delta_\mathrm{ij} / 2 m^* a^2$ with $m^*$ the effective mass of electrons and $\Delta_\mathrm{ij}$ the discrete two-dimensional Laplacian\cite{Datta1995}.
By defining the separate subsystems according to the electrostatic confinement, we apply the presented method and calculate the tunnel-coupling. To visualize the tunnel barrier in energetic height and width in 1D, we calculate a semiclassical tunneling path $l(x,y)$ of an electron. To calculate $l(x,y)$, we use the Dijkstra algorithm with on-site weights $\sqrt{2m^*a^2(V_\mathrm{ij}-\epsilon_\mathrm{S})/\hbar^2}$. These weights are motivated by the one dimensional WKB-approximation. Along this path the potential is evaluated and the tunnel barrier characterized. This is shown exemplary in \rfig{fig:combinedApp} (a) and for the discussed types of disorder in \rfig{fig:combinedApp} (e). Note that $l(x,y)$ is sensitive to numerical errors and is not used to calculate $t_\mathrm{C}$ by our ansatz. \newline
For the potential landscape of the SET without any disorder effect included, the tunnel couplings of the read-out QD with $\epsilon_\mathrm{S}=E_\mathrm{G}/2$ to the source reservoir is $t_\mathrm{L}^\mathrm{(ref)}=1.3 \ \mathrm{\mu eV}$. We obtained this result by using $\gamma_\mathrm{ext}=700 \ \mu \mathrm{eV}$ in \req{eq:Gmeff} and $N=36589$ sites. We computed $n=148$ eigenstates of system $\sysM + \sysL$ with energies in the vicinity of $\epsilon_\mathrm{S}$ and found quality indicators $f_\gamma=0.008$ and $f_n=21.16$. The determination of $n$ and the definition of the quality indicators is subject of the sections \rsec{Verification} and \ref{largeN}. The tunnel coupling to the drain reservoir is $t_\mathrm{R}^\mathrm{(ref)}=2.0 \ \mathrm{\mu eV}$ with $\gamma_\mathrm{ext}=700 \ \mu \mathrm{eV}$ and $N=36944$. We computed $n=148$ eigenstates with $f_\gamma=0.02$ and $f_n=19.19$. For deviations from the tunnel barrier potential maximum between \sysS and \sysM on the scale of the used spatial resolution $a$, we calculate an error of tunnel coupling $\Delta t_\mathrm{L}^\mathrm{(ref)} = 0.3 \ \mu \mathrm{eV}$ and $\Delta t_\mathrm{R}^\mathrm{(ref)} = 0.4 \ \mu \mathrm{eV}$.
The tunnel couplings $t_\mathrm{L}^\mathrm{(ref)}$ and $t_\mathrm{R}^\mathrm{(ref)}$ are used as reference values for the effect of different disorder types on tunnel couplings.
\subsection{Ge-Ge bond disorder}
In SiGe unit cells, the specific arrangement of Si and Ge atoms in the diamond lattice leads to energy variations of the conduction band edge. From tight-binding simulations of periodic SiGe unit cells, Ge-atoms on neighboring sites decrease the conduction band by approximately $\delta V = 100 \ \mathrm{meV}$ compared to a fully random barrier \cite{Jiang2012} and hence increase the energy of the electrons locally on the spatial resolution of an 8-atom unit cell.
To model this disorder effect, we assume a binomial distribution $p_n(x)$ to find $n$ Ge-Ge bonds  surrounding a Ge-occupied site given an alloy composition factor $x$. To weight the disorder effect with respect to the electron envelope wavefunction $\Psi(z)$, the wavefunction overlap $F=\int_{z_\mathrm{I}}^\infty |\Psi(z)|^2 dz $ with the SiGe layer ($ z > z_\mathrm{I} $) is included, where $z_\mathrm{I}$ is defined at the Si/SiGe interface. The resulting distribution and magnitude of potential variations $\Delta V$ over the number of Ge-Ge bonds $n$ surrounding a single atom is
\begin{equation}
\begin{aligned}
\mathrm{dist}( \Delta V  ) &= p_n(x) x^n (1-x)^{4-n} \ \mathrm{and} \\
\mathrm{magn}( \Delta V  ) &= -\dfrac{n}{2} \delta V  F \ \mathrm{.}
\end{aligned}
\end{equation} 
The factor $1/2$ in $\mathrm{magn}( \Delta V )$ accounts for double counting of each bond when iterating over the 8-atom unit cell. Finally, we define the length scale of this fluctuation by the lattice constant of the $\mathrm{Si}_{(1-\mathrm{x})} \mathrm{Ge}_\mathrm{x}$-alloy with $\lambda_\mathrm{Ge-Ge \ bonds} \approx 0.5 \ \mathrm{nm}$ and transfer the presented potential variations to our tight-binding model.\newline
The model results in a number of Ge-Ge bonds $n = 6\pm 4$ per $\mathrm{nm}^2$, where we neglect further variations along z. The non-zero average of Ge-Ge bonds leads to an average increase of the electron energy of $\Delta \bar{V} = 2.5  \ \mathrm{meV}$. This energy offset is neglected within the following study, since it is compensated by an initial tuning of the electrostatic potential.
By adding $\Delta V$ to the electrostatic potential, the tunnel-coupling can be calculated as before. The resulting effect on the electrostatic potential is shown by the semiclassical tunnel path in \rfig{fig:combinedApp} (e). The normalized distribution $\sigma(t_\mathrm{C})$ of the calculated tunnel-coupling $t_\mathrm{C}$ for $N_\mathrm{dis}=10^4$ randomly generated Ge-Ge bond ensembles is shown in \rfig{fig:combinedApp} (d).
Due to the small length scale $\lambda_\mathrm{Ge-Ge \ bonds}$ and the comparable magnitude of the variation with respect to the barrier height, this modeled disorder leads to varying tunnel-couplings within two orders of magnitude compared to the reference value.

\subsection{QW step disorder}
Interface roughness has been reported to be a major source of disorder leading to variations of the valley splitting\cite{Kharche2007a}. Furthermore, atomic steps at the interface of Si/SiGe result in changes of the confinement along the growth direction and hence to a fluctuation in the energy of the electrons.
To model this effect, we restrict to relative changes of only one step at each interface. Assuming effective single-layer growth using molecular beam epitaxy (MBE), the step height is $h_\mathrm{Step}=a_\mathrm{SiGe}/4=0.135 \ \mathrm{nm}$. This leads to three different confinement energies $E_{0}$, $E_+$ and $E_-$ along z. $E_{0}$ is the energy for a QW without any additional step. $E_+$ is the energy for a QW with a width decreased by one interface step $h_\mathrm{Step}$ and $E_-$ for a QW with a width increased by $h_\mathrm{Step}$.
The resulting potential variation is 
\begin{equation}
\Delta V_\pm = - (E_{0} - E_{\pm} ) ,
\end{equation} 
where $E_{0}$ and $E_{\pm}$ are the energies of the three different confinements as defined above.
These energies are calculated for an applied voltage bias of $E_\mathrm{G}/e$, a QW width of $12$ nm and a conduction band minima difference of $\Delta E_c=160$ meV of the heterostructure. There is no potential offset $\Delta \bar{V}_\pm=0$ by construction. Furthermore, we define the length scale of this fluctuation to vary uniformly in the range of $\lambda_\mathrm{QD step}= 1 \ \dots \ 24 \ \mathrm{nm}$ corresponding to wafer miscut angles of $\alpha = 7.8\degree \dots 0.3 \degree $. The tunnel-coupling is calculated as before. The effect of this type of disorder on the semiclassical tunnel path is shown in \rfig{fig:combinedApp}(e), where due to the small magnitude of approximately hundred $\mathrm{\mu eV}$, the potential fluctuations are multiplied by a factor of $10$. Due to the relatively long coherence length $\lambda_\mathrm{QW step}$ and the small magnitude of the variation compared to the tunnel barrier height, this disorder effect leads to variation in the tunnel-couplings smaller than one order of magnitude compared to the reference value as shown in \rfig{fig:combinedApp} (d).

\subsection{Impurities}
\begin{figure}[htb]
\begin{center}
	  \includegraphics[width=0.48\textwidth]{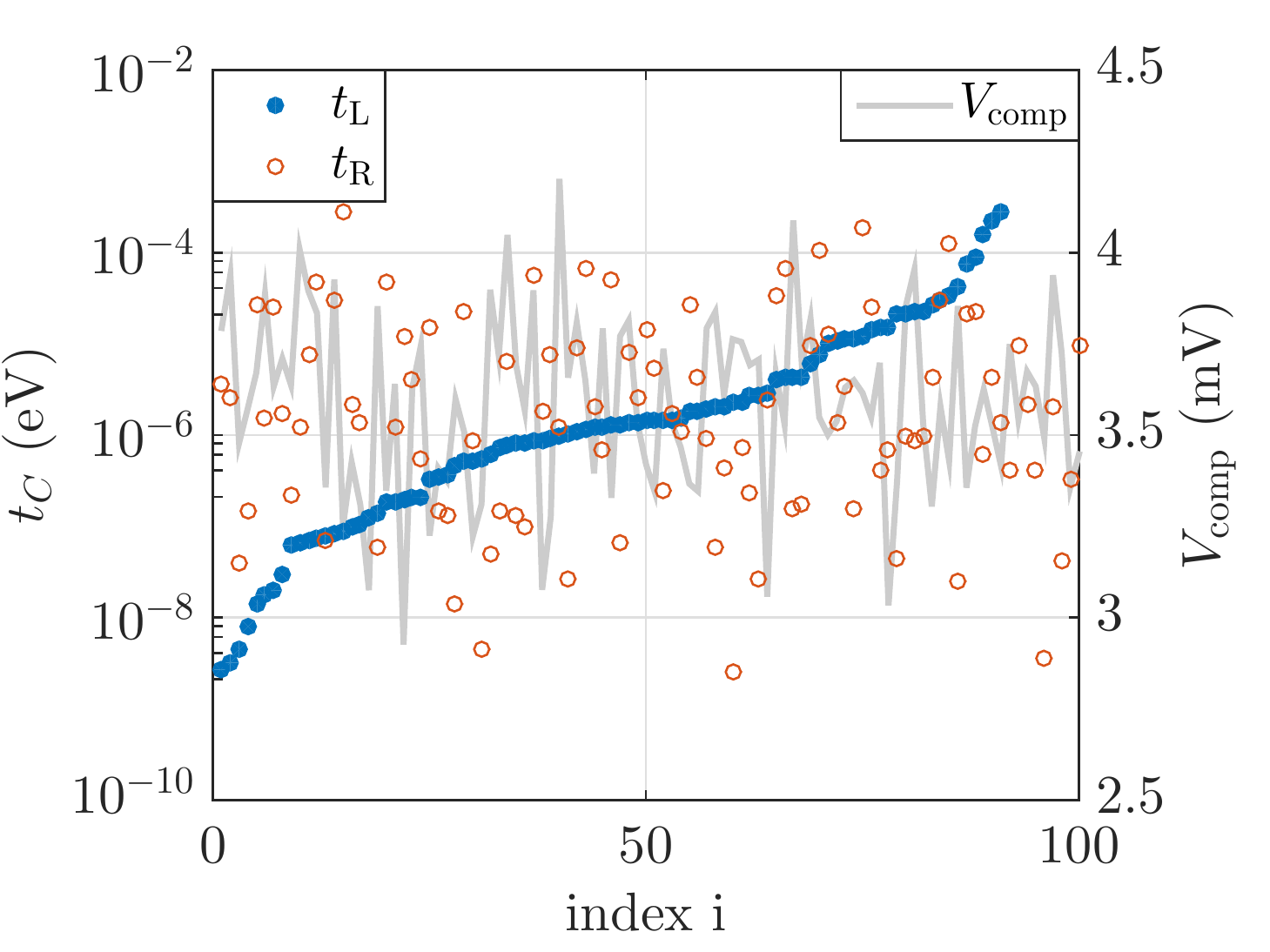}
\end{center}
\caption{Calculated tunnel-coupling for $N_\mathrm{dis}=100$ randomly distributed positively charged impurities with a concentration of $10^{10} \ \mathrm{cm}^{-2}$. The presented data is sorted with respect to $t_\mathrm{L}$. For every impurity distribution $i$, the resulting tunnel-couplings $t_\mathrm{L}$, $t_\mathrm{R}$ and the corresponding global compensation voltage $V_\mathrm{comp}$ are shown. 
}
\label{fig:imputitytC}
\end{figure}

We refer to positively charged defects located in the heterostructure as impurities. Impurities formed by oxygen atoms located near and within the Si QW have been reported with concentrations of $10^{10}\dots 10^{11} \ \mathrm{cm}^{-2}$ introduced during the growth of the heterostructure in a chemical vapor deposition reactor (CVD) \cite{Mi2015a}. 
Remote impurities located at the interface between the heterostructure and an \AlO oxide layer have been suggested to dominate electron scattering\cite{Li2013}.\newline
In this section, we introduce impurities located at the interface of the heterostructure and a possible oxide layer $34 \ \mathrm{nm}$ above the QW. All impurities are positively charged with $q_\mathrm{Imp}=-e$ with $e$ the electron charge and randomly distributed over the interface leading to a concentration of $10^{10} \ \mathrm{cm}^{-2}$. \newline
In comparison to the initially tuned potential without disorder, these impurities lead to a resulting average positive offset $\bar{V}_\mathrm{Imp} \approx 3.5 \ \mathrm{meV}$ and potential fluctuations of $\Delta V_\mathrm{Imp} \approx 5 \ \mathrm{meV}$ as shown in \rfig{fig:combinedApp}(c). 
%Since the typical length scale depends on the position of the impurities, we define $\lambda_\mathrm{Imp}\approx 50 \ \mathrm{nm}$ as the diameter of the induced electrostatic potential of a single charge, located $34 \ \mathrm{nm}$ above the QW, exceeding $1 \ \mathrm{meV}$.
We compensate $\bar{V}_\mathrm{Imp}$ by a global voltage offset of $V_\mathrm{comp}=\bar{V}_\mathrm{Imp}$ applied to all used gates for each single impurity ensemble. This is a rather simple compensation scheme which only requires a global voltage parameter to be set. In this manner, we compensate the potential and end up with two tunnel barriers with a probability of $59\%$ and at least one tunnel barrier with a probability of $96\%$ using 100 randomly chosen impurity ensembles. Within an experiment, the global voltage could be tuned more precisely to achieve the desired tunnel couplings.\newline
By calculating the tunnel-coupling from the source reservoir into the read-out QD (left barrier) and the read-out QD into the drain reservoir (right barrier), we quantify the effect of impurities on the functionality of our SET for several different impurity distributions. The resulting distribution of the tunnel-couplings is shown in \rfig{fig:combinedApp} (d) and varies over four orders of magnitude using the simple compensation scheme. This type of disorder resembles the strongest variation in $t_\mathrm{C}$ compared to the previous discussed effects. We observe differences in the tunnel-couplings between the left and the right tunnel barrier up to several  $\mathrm{meV}$, as shown in \rfig{fig:imputitytC}. The different distributions of positively charged impurities are indexed by $i$ and sorted with respect to the tunnel-coupling of the left barrier. Ensembles with only one remaining tunnel barrier are included within this plot ($i \ge 91$) and show the largest disorder impact on the tunnel barriers. Here the Fermi energy exceeds the height of the left tunnel barrier. Note that left and right tunnel barriers are uncorrelated in \rfig{fig:imputitytC}. Thus, a precisely tuned global voltage is insufficient to tune both tunnel couplings. It requires involved individual tuning of gate voltages to set both tunnel barriers as desired.\newline
\section{Conclusion}\label{Conclusion}
We present a method for calculating tunnel-couplings of open quantum systems. We aim especially at the simulation of gate patterns and disorder for gated semiconductor quantum computers. We apply this method to a gate layout of an SET charge detector as a showcase.
The method is applicable to various systems and is flexible with respect to available computational resources while including all modeled details of the electrostatic potential. The Markov approximation is solely used for the reservoir region far away from the barrier. Basic models for three different disorder sources, typical for Si/SiGe heterostructures, are used to study the effect of electrostatic disorder on the tunnel-coupling of the SET, pointing towards charged defects as a strong source of varying tunnel-coupling over four orders of magnitude. While a detailed model of disorder potential in Si/SiGe is beyond the scope of our work, we expect that our method can be used to calculate tunnel-couplings with improved noise models of various material systems. 

\section*{Acknowledgments}
We thank F. Haupt for proof-reading the manuscript. P. Cerfontaine acknowledges support by Deutsche Telekom Stiftung.

\bibliographystyle{apsrev4-1}
\begin{appendices}\label{Appendix}
\section{Validation}\label{Verification}
To test this approach, we apply the presented method on an analytically solvable tight-binding toy-model system. The validation focuses on the use of the presented Markov approximation and on the calculated tunnel coupling.
The toy-model system is two dimensional and consists of a single site coupled by the transition element $w$ to a 2D lattice with $N$ sites. Within the 2D lattice, adjacent sites are coupled by nearest-neighbor transition elements $t$. To define the presented subsystems, the 2D toy-model is schematically shown in \rfig{fig:combinedVer}(a). The single site is defined as subsystem \sysS depicted in blue. Subsystem \sysM is defined by all inner sites of the 2D lattice depicted in red. Without further restriction subsystem \sysS is coupled to the middle site of subsystem \sysM. The outer sites of the 2D lattice are defined as subsystem \sysL depicted in yellow. Since we approximate the lead system, it is sufficient to only account for the sites, which are directly coupled to system \sysM.\newline

\subsubsection*{Markov approximation}
First, we validate the used Markov approximation in the lead system. Therefore, we compare the computed numerical density of states of the 2D toy-model system using the Markov approximation with the density of states for a discrete infinite 2D lattice.
The latter is calculated analytically\cite{Piasecki} :
\begin{equation}\label{eq:2DAnal}
\begin{aligned}
\rho_\mathrm{M,analytic}(\hbar\omega) &= \dfrac{1}{2t\pi^2}  K \bigg ( 1-\bigg (\dfrac{\hbar\omega-V_\mathrm{0}-4t}{4t} \bigg )^2 \bigg ) \\
\mathrm{for} \ 0 &< |\hbar\omega| \leq V_\mathrm{0} + 4t \ .
\end{aligned}
\end{equation}
Note that if the analytic expression of $\rho_\mathrm{M,analytic}$ were known for all problems, we could use $\rho_\mathrm{M,analytic}(\hbar\omega)$ in \req{eq:Grscalar} and calculate $H_\mathrm{M,eff}$ by \req{eq:Hmeff} and thus $t_\mathrm{C}$ analytically. Since in most realistic problems $\rho_\mathrm{M,analytic}$ is unknown, we use \req{eq:Markov}, the Markov approximation instead of the Kramers-Kronig relation in \req{eq:Grscalar}.\newline
%In the following, we use the Markov approximation defined by $G_\mathrm{L}(\hbar\omega) = -i/2t\pi$.\newline
For a 2D toy-model with $N=1521$ sites and $t=1$, the numerical and analytic density of states are shown in \rfig{fig:combinedVer}(b).
The numerical density of states of the 2D lattice can be calculated using \req{eq:Gmeff} with $\rho_\mathrm{M}(\hbar\omega)=-i \mathrm{Tr}[ \hat{G}_\mathrm{M}(\hbar\omega) ]$\cite{Datta1995}.
Up to a fluctuation of the numerical density of states, both solutions coincide and follow the same behavior with respect to the energy $\hbar \omega$. For $\hbar\omega\approx 0$, both solutions exhibit a van-Hove-singularity\cite{Hovz1952}. 
In the vicinity of the energy band edge $|\hbar\omega| \approx 4t$, the deviation between the analytic and numerical solutions increases. This is explained by a decreasing imaginary part of the energy-levels, leading to more $\delta$-function shaped states. 
For energies $0<|\hbar\omega|<4t$, the analytic density of states is rather constant. In comparison to the infinite system, the finite size of the model leads to an overall fluctuation. Focusing on states $|\Psi_m \rangle$ and calculating $\mathrm{Re}(\langle \Psi_m | \hat{G}_\mathrm{M}(\hbar\omega)) | \Psi_m\rangle$, all energy-levels are approximately Cauchy-Lorentz shaped. Due to the non-equidistant energetic distribution of energy levels, the energetic overlap of neighboring states varies resulting in a non-constant density of states (see inset in \rfig{fig:combinedVer}(a)). Hence, the fluctuation is a function of energy $\hbar\omega$ and system size $N$.\newline
For a finite number of sites $N$ in the 2D toy-model, this fluctuation can be compensated by an additional external decay parameter $\gamma_\mathrm{ext}$ used as regularization factor, which is added to $i\gamma_m \rightarrow i(\gamma_m+\gamma_\mathrm{ext})$ as already introduced in \req{eq:Gmeff}. To define a quality indicator for the fluctuation, we use 
\begin{equation}\label{Measure1}
f_\gamma = \dfrac{A_\mathrm{\rho_\mathrm{M}}}{\rho_\mathrm{M, max}}\bigg \vert_{\hbar\omega=\epsilon_\mathrm{S}} \ \mathrm{,}
\end{equation}
where $A_\mathrm{\rho_\mathrm{M}}(\hbar\omega)$ is the maximum amplitude of the local fluctuation defined on the energetic range of multiple neighboring states and $\rho_\mathrm{M, max}$ is the maximal value of $\rho_\mathrm{M}$ both evaluated at the same energy $\hbar\omega$ (see inset of \rfig{fig:combinedVer}(b)). 
To compensate the fluctuation, we increase $\gamma_\mathrm{ext}$ until the $\partial f_\rho / \partial \gamma_\mathrm{ext}$ saturates at a minimum. In this way, we determine to optimal value for $\gamma_\mathrm{ext}$ labeled $\gamma_\mathrm{ext,opt}$. For the validation of the toy-model and the numeric calculation of the tunnel couplings of the SET, we used $A_\mathrm{\rho_\mathrm{M}}(\hbar\omega)$ on the energetic interval of five neighboring energy-levels after subtracting the overall tendency of $\rho_\mathrm{M}$ approximated by a linear offset.\newline
Determining $\gamma_\mathrm{ext,opt}$ as described above, the numerical and analytic density of states coincide very well for the used energy and a given number of sites $N$ as can be seen from \rfig{fig:combinedVer}(c) for $\hbar\omega= -2$ with all states $N$ taken into account. Note that particularly for the model validation, $f_\gamma$ does not include any information of the analytic solution.\newline

\subsubsection*{Tunnel coupling}
Now, we validate the calculation of the tunnel coupling $t_\mathrm{C}$ using the presented method on our 2D toy-model.
By using a constant on-site potential $V_\mathrm{0}$ in system $M+L$, we calculate the analytic solution of the tunnel coupling for an arbitrary energy $\hbar\omega$ to be
\begin{equation}
    t_\mathrm{C,analytic}(\hbar\omega) = 2\pi w^2 \rho_\mathrm{M,analytic}(\hbar\omega)
\end{equation}
with $w=0.1$ assuring weak-coupling of \sysS to \sysM+\sysL.
In the following, we explicitly focus on a single energy-level $|\mathrm{s}\rangle$ in subsystem \sysS with energy $\epsilon_\mathrm{S}$ and energy-conserving tunneling.\newline
The computed relative tunnel couplings $t_\mathrm{C}/t_\mathrm{C,analytic}$ are shown for the two dependencies $\hbar\omega=\epsilon_\mathrm{s}$ and $N$ in \rfig{fig:combinedVer}(d). For every point, $f_\rho$ is minimized by $\gamma_\mathrm{ext,opt}$.\newline
By varying $\epsilon_\mathrm{s}$, the numerical tunnel coupling differs from the analytic solution with an error of up to $3\%$ where $\gamma_\mathrm{ext,opt}$ does not exhibit a clear tendency over $\hbar\omega=\epsilon_\mathrm{s}$ and only varies due to the varying local fluctuation.
For different system sizes $N$, the numerical tunnel coupling differs from the analytic solution by an error of up to $6\%$ where $\gamma_\mathrm{ext,opt}$ decreases for increasing system sizes $N$. Since in a tight-binding model with system size $N$, there are exactly $N$ energy-levels, the intrinsic energetic difference between neighboring states decreases with increasing system size, resulting in a decreasing $\gamma_\mathrm{ext,opt}$.
With system size $N>10^3$, the error may be reduced even further, but might lead to computational challenges.
\newline

\section{Transfer to large system sizes}\label{largeN}
\begin{figure*}[htb]
\begin{center}
    	  \includegraphics[width=0.99\textwidth]{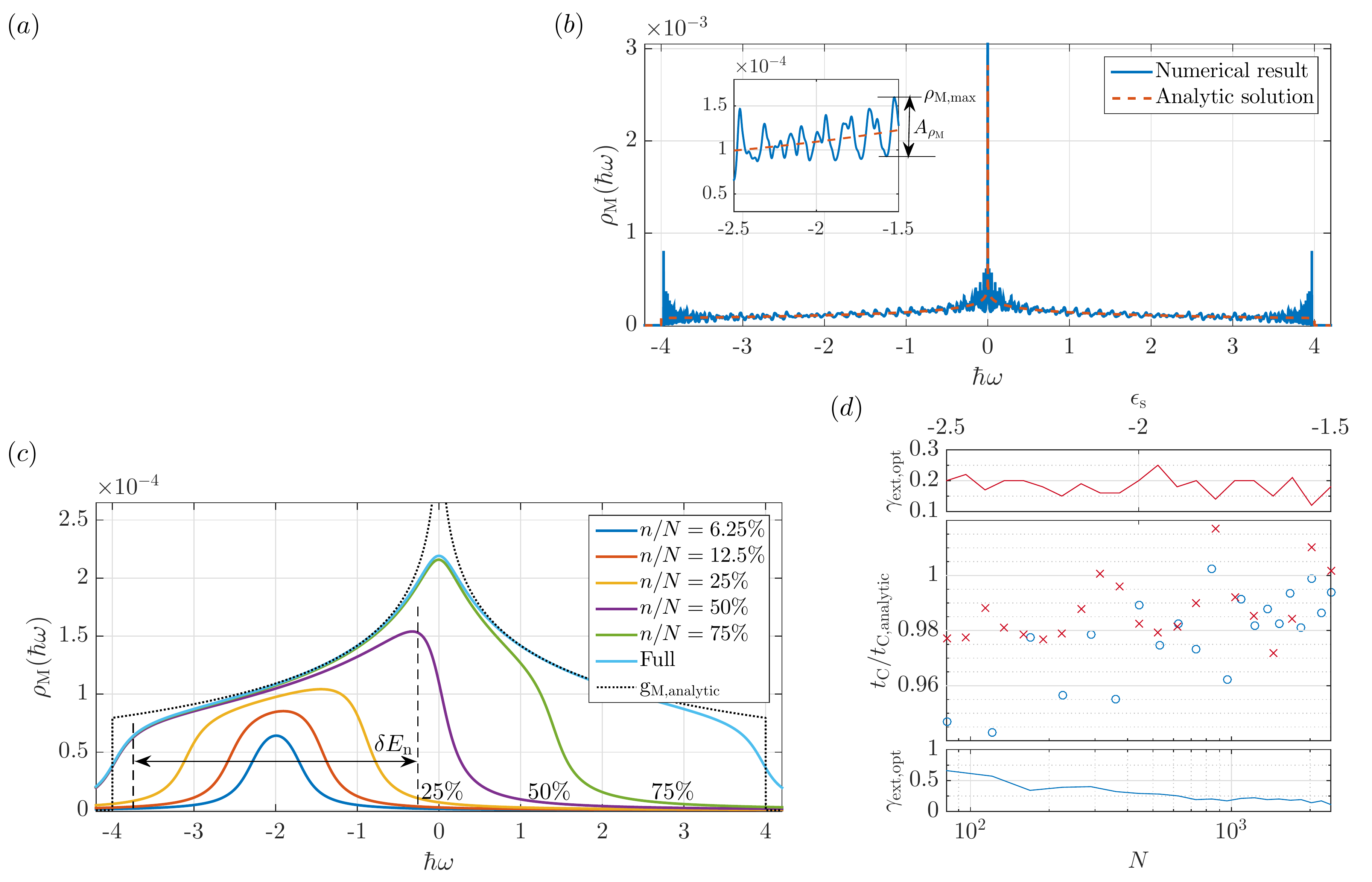}
            \llap{\makebox[0.96\textwidth][l]{\raisebox{7.1cm}{ 
                    \includegraphics[width=0.35\linewidth]{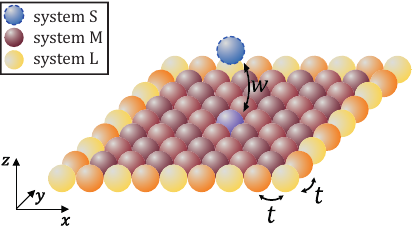}
		        }}}
\end{center}
\caption{
(a) Schematic of the 2D tight-binding \toymodel with a single site coupled by the transition element $w$ to a 2D lattice with $N$ sites, each coupled by the nearest-neighbor transition element $t$. 
(b) Comparison of the density of states for an infinite 2D lattice: numerical solution using Markov approximation vs. analytic solution. For $\hbar\omega\approx 0$, both solutions diverge due the a van-Hove singularity. Close to the band edge with $|\hbar\omega| \approx 4t$, the deviation between the analytic and numerical solutions increases. In the numerical simulation, the 2D lattice consists of $N=1521$ sites coupled by nearest-neighbor coupling with $t=1$. 
(c) Density of states for an 2D lattice with $j_\mathrm{ext,opt}=0.2$ chosen with respect to $N=1521$ and $\hbar\omega=\epsilon_\mathrm{S}=-2$ for different calculated fractions $n/N$ of the full solution of $H_\mathrm{M}$.  By reducing the fraction of computed states $n/N$, $\rho_{\mathrm{M},n<N}$ deviates from theory.
(d) Dependency of the numerical tunnel coupling $t_\mathrm{C}$ and the optimal external decay $\gamma_\mathrm{ext,opt}$ on system size $N$ of the 2D lattice and energy-level $\epsilon_\mathrm{S}$ of the single site. The single site is coupled by the transition element $w=0.1$. The 2D lattice is defined with $N=1521$ and $t=1$. Varying $\epsilon_\mathrm{S}$, the numerical tunnel couplings differs from the analytic solution up to an error of $3\%$, where $\gamma_\mathrm{ext,opt}$ shows a constant tendency.
Varying $N$, the numerical tunnel couplings differs from the analytic solution up to an error of $6\%$. The external decay $\gamma_\mathrm{ext,opt}$ shows a decreasing tendency for increasing $N$.
}
\label{fig:combinedVer}
\end{figure*}

For large system sizes e.g. in our realistic showcase with $N\approx 10^5$, solving the Schrödinger equation of the complete system may exceed available computational resources. Therefore, we discuss the influence of computing only $n$ states around $\epsilon_\mathrm{S}$ of the total $N$ states in subsystem \sysM. The numerical density of states for different fractions $0<n/N \le 1$ is shown in \rfig{fig:combinedVer}(c).
For the full solution of the Schrödinger equation ($n=N$), the analytic and numerical density of states coincide using $\gamma_\mathrm{ext,opt}=0.2$ for $\epsilon_\mathrm{S}=-2$ and $  N=1521$. Due to the external decay, the van-Hove singularity at $\hbar\omega=0$ is suppressed and the band edges at $|\hbar\omega|=4t$ are smeared out. Since the overlap of energetically far distant states is negligible, we only compute states within the energetic proximity of $\epsilon_\mathrm{s}$. This reduces the required computational resources drastically. For lower fractions (here $n/N<50\%$), the numerical density of states deviates from the analytic solution since we neglect states which contribute to the density of states and tunnel coupling at the energy $\epsilon_\mathrm{s}$.
Similar to $f_\rho$, we define an additional quality indicator:
\begin{equation}\label{Measure2}
f_n = \dfrac{\delta E_n}{\gamma_\mathrm{m} +\gamma_\mathrm{ext}} \ \mathrm{,}
\end{equation}
where we use the energetic interval $\delta E_n$ which is spanned by these $n$ computed states with respect to the broadening $\gamma_\mathrm{m} +\gamma_\mathrm{ext}$ of the states in close energetic proximity to $\epsilon_\mathrm{s}$. This is shown for $n/N=50 \% $ in \rfig{fig:combinedVer}(c).
\newline
Within the validation, we explicitly focused on small system sizes ($N \approx 10^3$) leading especially to errors due to finite sizes of the system. For larger system sizes with $N \gtrsim 10^5$ sites, this error is captured intrinsically and small additional external decay $\gamma_\mathrm{ext}$ can be included to minimize $f_\rho$. On the other hand, the second indicator in \req{Measure2} leads to a tremendous reduction of computational resources dominated by the dimension of $H_\mathrm{M,eff}$, while still assuring reasonable results.

\end{appendices}

\end{document}